\documentclass[12pt]{iopart}
\usepackage{graphicx}
\newcommand{\Vec}[1]{\mbox{\boldmath$#1$}}
\begin{document}
\title{Theoretical Aspects of  Charge Correlations in
$\theta$-(BEDT-TTF)$_2X$}
\author{
Kazuhiko Kuroki
}
\address{
Department of Applied Physics and
Chemistry, The University of Electro-Communications,
Chofu, Tokyo 182-8585, Japan}
\begin{abstract}
A review is given on the theoretical studies of 
charge correlations in  $\theta$-(BEDT-TTF)$_2X$. 
Various studies show that within a purely electronic model on 
the $\theta$-type lattice with the on-site $U$ and the  
nearest neighbor $V_p$ and $V_c$ interactions, the diagonal stripe, 
$c$-axis three-fold, and the vertical stripe charge correlations are 
favored in the regime $V_p< V_c$, $V_p\sim V_c$, and $V_p> V_c$, 
respectively. In the realistic parameter regime of $V_p\sim V_c$, 
there is a competition between $c$-axis three fold state and the 
diagonal stripe state. Since these are different from the 
experimentally observed $a$-axis three fold and the horizontal stripe 
charge correlations, additional effects have to be included in order to 
understand the experiments. The electron-lattice coupling, 
which tends to distort the lattice into the $\theta_d$-type,  
is found to favor the horizontal stripe state, 
suggesting that the occurrence of this stripe ordering 
in the actual materials may not be of purely electronic origin. 
On the other hand, distant electron-electron 
interactions have to be considered in order to understand 
the $a$-axis three fold correlation, whose wave vector is close to the 
nesting vector of the Fermi surface. These studies seem to suggest that the 
minimal model to understand the charge correlation in 
$\theta$-(BEDT-TTF)$_2X$ may be more complicated than expected. 
Future problems regarding the 
competition between different types of charge correlations are discussed.
\end{abstract}


\section{Introduction}

\begin{figure}
\begin{center}
\includegraphics[width=8cm]{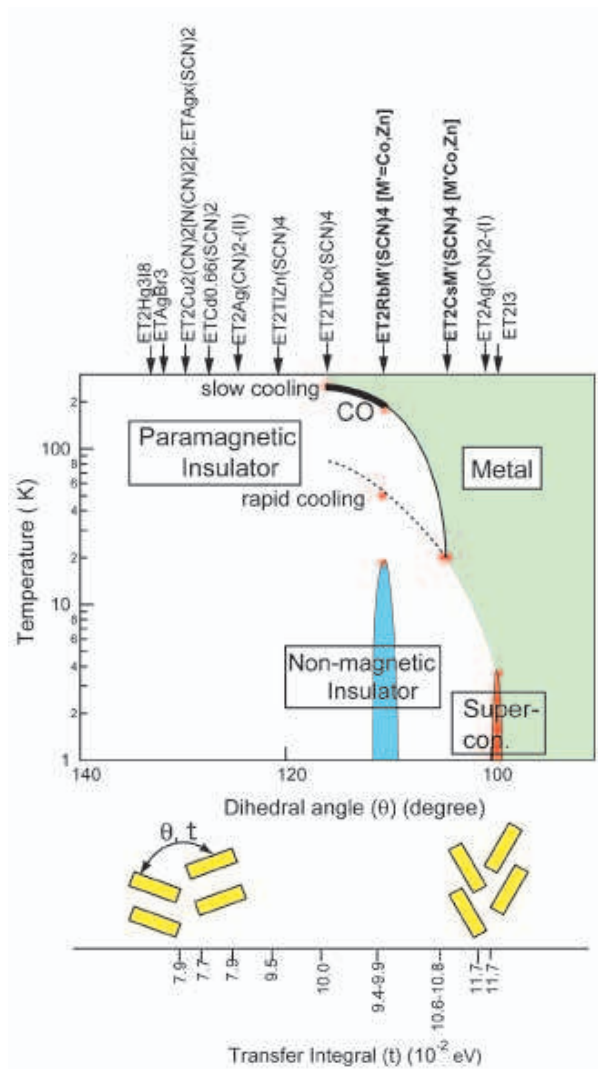}
\caption{Mori's phase diagram for $\theta$-(BEDT-TTF)$_2 X$.  
Reproduced with permission from ref. \cite{Morirev}. 
Copyright 2006 by the Physical Society of Japan.}
\label{fig1}
\end{center}
\end{figure}

\begin{figure}
\begin{center}
\includegraphics[width=8cm]{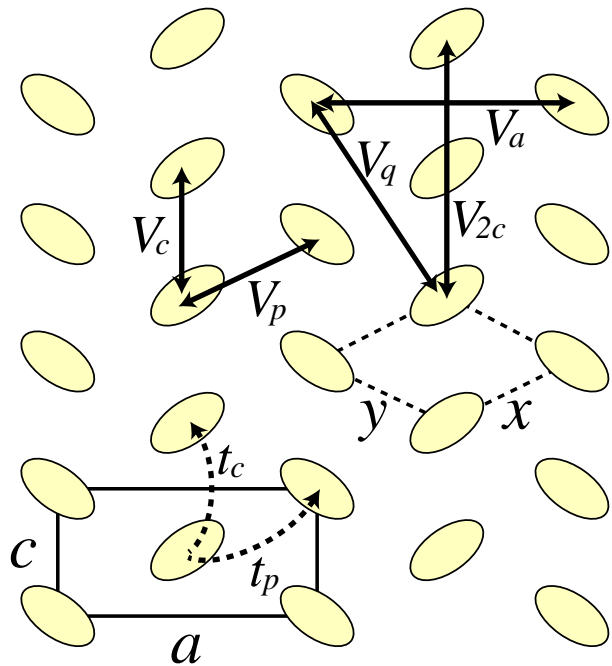}
\caption{Lattice structure of the cation layer of 
$\theta$-(BEDT-TTF)$_2 X$. $t_p$ and $t_c$ are the nearest neighbor hopping 
integrals. $(t_p,t_c)=(106,-5)$, $(137,-49)$, and $(99,-33)$ [meV] 
for CsCo, CsCo under pressure (10 kbar), and RbCo salts, respectively,
according to the extended H\"{u}ckel estimation \cite{Watanabe1999}. 
$V_p$ and  $V_c$ are the nearest-neighbor 
interactions, while $V_a$, $V_q$, and $V_{2c}$ are the next-nearest-neighbor 
interactions. The $a$-$c$ unit cell is the usual unit cell, while 
we can use the $x$-$y$ unit cell to unfold the Brillouin zone.}
\label{fig2}
\end{center}
\end{figure}

\begin{figure}
\begin{center}
\includegraphics[width=5cm]{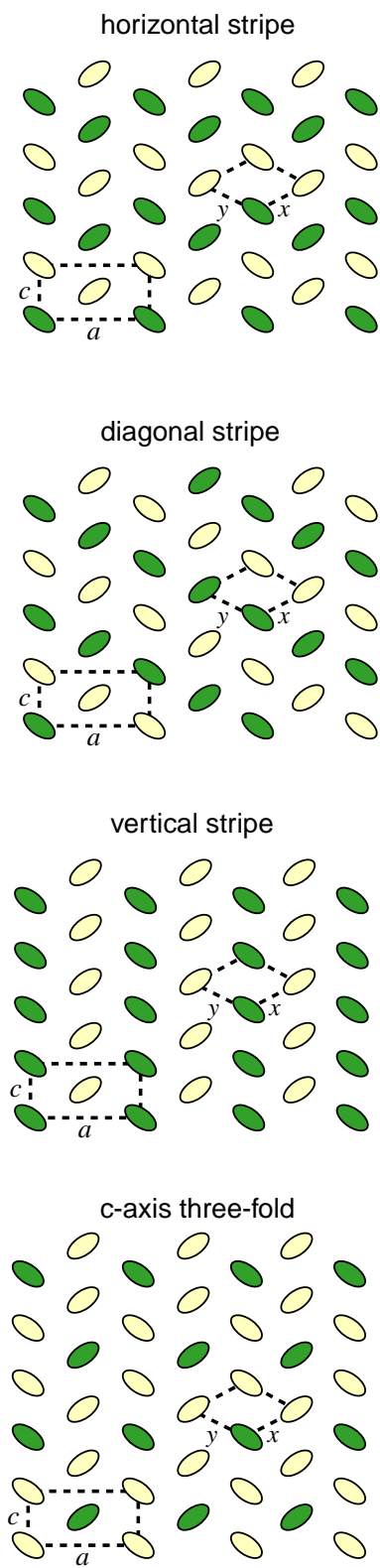}
\caption{Horizontal stripe, 
vertical stripe, diagonal stripe, and 
$c$-axis threefold charge correlations.
The green (or dark) molecules are the charge-rich ones.}
\label{fig3}
\end{center}
\end{figure}

\begin{figure}
\begin{center}
\includegraphics[width=8cm]{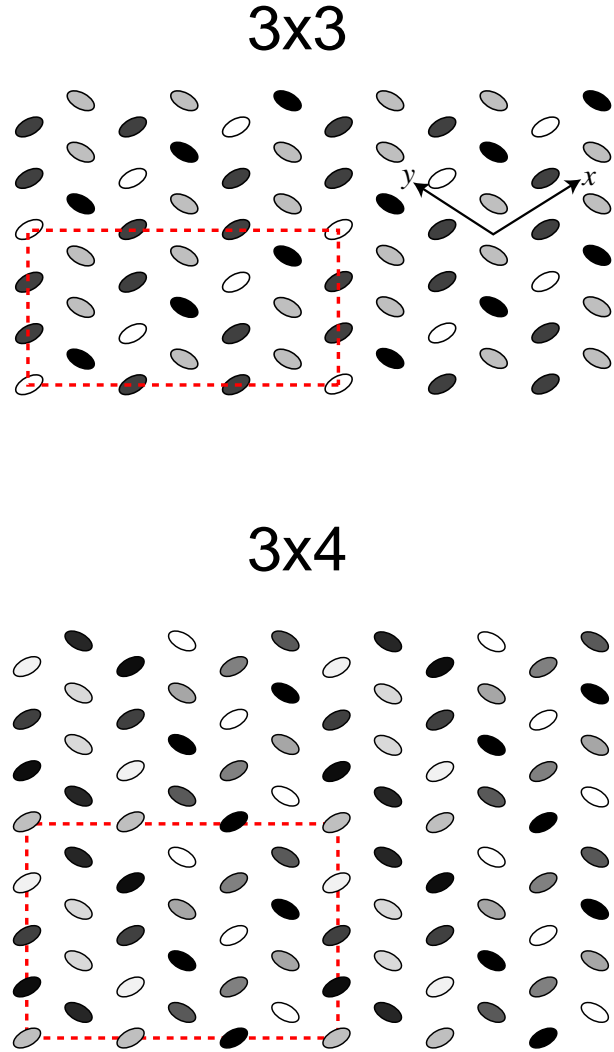}
\caption{$a$-axis threefold ($3\times 3$ and $3\times 4$) 
charge correlations.
The dark molecules are the charge-rich ones.}
\label{fig4}
\end{center}
\end{figure}

\begin{figure}
\begin{center} 
\includegraphics[width=8cm]{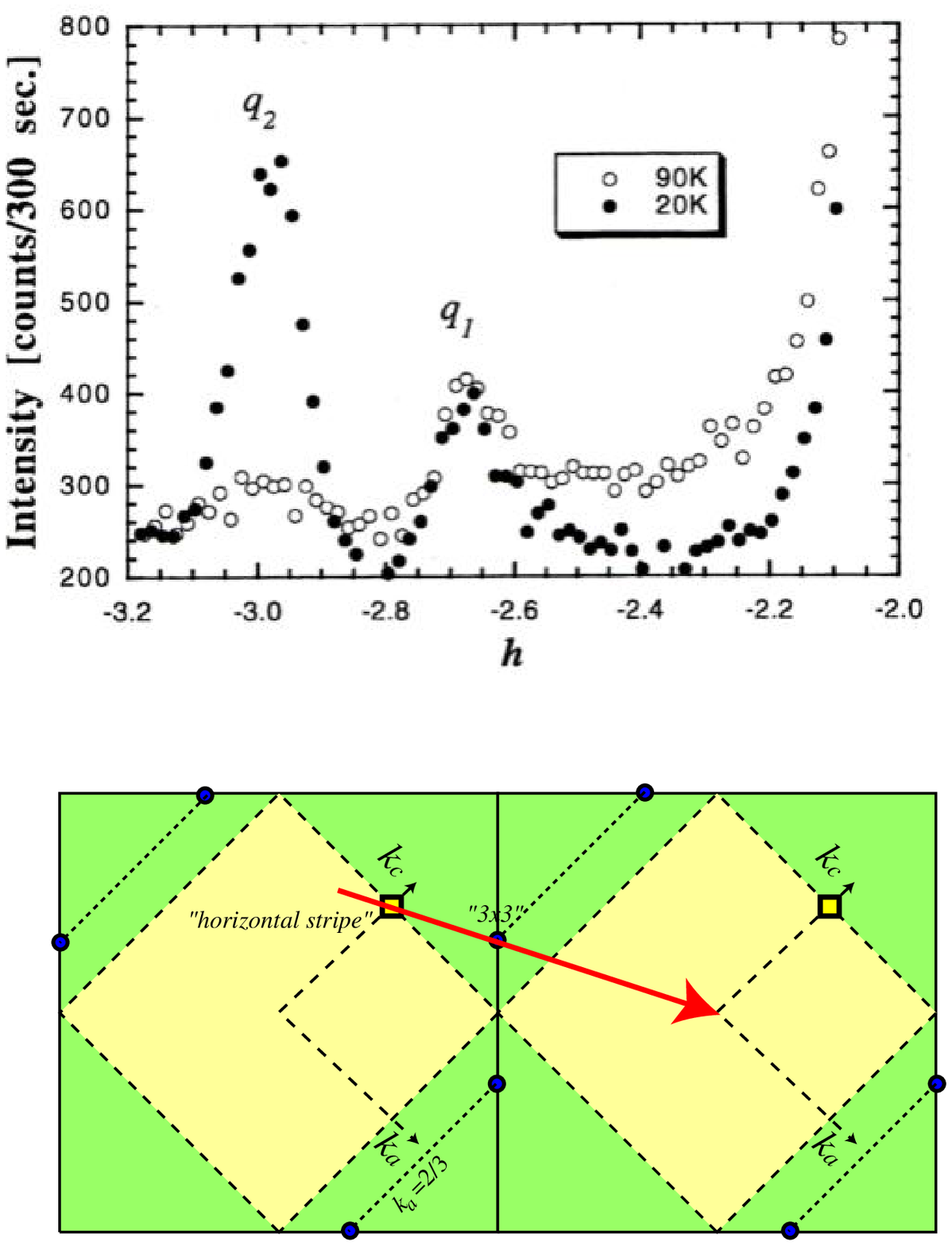}
\caption{Upper panel: X-ray scattering intensity scanned along a 
diffuse sheet, expressed in units of the reciprocal lattice primitive vectors 
as $(\bar{3}\:\:  1\:\: 1/2)+(2\xi\:\: 0\:\: \xi)$ 
with $-0.09\leq \xi \leq 0.48$, 
where the horizontal axis in the figure is $h=-3+2\xi$. 
$q_1$ is the wave vector corresponding to the 
$a$-axis threefold ($3\times 3$) correlation, while 
$q_2$ corresponds to the horizontal stripe correlation. 
Reproduced with permission from ref. \cite{Watanabe1999}. 
Copyright 1999 by the 
Physical Society of Japan. Lower panel: 
line (the arrow) along which the X-ray scan is taken (i.e., the diffuse sheet) 
shown in the Brillouin zone adopted in theoretical studies 
(see figure \ref{fig7}).}
\label{fig5}
\end{center}
\end{figure}

In recent years, charge correlations, i.e., 
orderings and fluctuations have been one of the central 
issues in the physics of organic materials
\cite{Takahashirev,Seorev,commentorder}.
$\theta$-(BEDT-TTF)$_2 X$\cite{class} is 
one of the most interesting families of organic compounds that 
exhibit short- or long-range charge correlations.
The crystal structure of $\theta$-(BEDT-TTF)$_2 X$, 
where BEDT-TTF = bis(ethylenedithio)tetrathiafulvalene, consists 
of a stack of alternating layers of BEDT-TTF and the anion $X$. In the 
BEDT-TTF layers, the molecules form an anisotropic triangular lattice,  
and the anion $X$ controls the angle between the BEDT-TTF molecules, 
which in turn determines the band structure. 
Mori and coworkers \cite{HMori1,HMori2} obtained the electronic phase 
diagram in the angle-temperature space as shown in figure \ref{fig1},
\cite{Morirev} which shows the 
wide variety of this series of materials, ranging from charge-ordered 
insulators such as $X=$RbZn(SCN)$_4$ to the superconductor $X=$I$_3$.
Further interest in this series of compounds has arisen by the 
observation of giant nonlinear transport for 
$X=$Cs$M'$(SCN)$_4$ ($M'=$Co,\ Zn), which causes the material 
to act as an organic thyristor \cite{Inagaki,Sawano}.
There, it has been pointed out that the coexistence of two 
types of short range charge correlations with modulation wave vectors 
$(q_a,q_b,q_c)=(\frac{2}{3},k,\frac{1}{3})$ and $(0,k,\frac{1}{2})$
(in units of the reciprocal lattice primitive vectors in the $a$, $b$, 
and $c$ directions; the $a$ and $c$ directions are shown in figure \ref{fig2}) 
plays an important role in this phenomenon.
The $(0,k,\frac{1}{2})$ charge modulation 
is often referred to as horizontal stripe ordering 
since this wave vector corresponds to a stripe-type charge modulation, 
where the stripes run 
in the horizontal direction (figure \ref{fig3}). 
The $(\frac{2}{3},k,\frac{1}{3})$ modulation 
corresponds to a $3\times 3$ charge correlation with respect to the 
$a$-$c$ unit cell, as shown in figure \ref{fig4}. 
At high temperatures,
a diffuse X-ray spot is observed at $(\frac{2}{3},k,\frac{1}{3})$
\cite{Nogami,Watanabe1999}, while the $(0,k,\frac{1}{2})$ structure develops 
as the temperature is lowered, and the system becomes more insulating 
(figure \ref{fig5}).
When an electric field is applied, 
the $(0,k,\frac{1}{2})$ horizontal stripe ordering is degraded, 
resulting in a recovery of the metallic behavior and thus the 
nonlinear transport.
Related to this phenomenon is the observation of lattice modulation with 
wave vector $(\frac{2}{3},k,0.29-\frac{1}{3})$  
for $X=$CsCo(SCN)$_4$ at a pressure of 10 kbar,
which has been attributed to a pressure induced $2k_F$ 
charge density wave  because the modulation wave vector 
coincides with the nesting vector of the Fermi surface \cite{Watanabe1999}. 

For $X=$Rb$M'$(SCN)$_4$ ($M'=$Co,Zn) 
also, two types of charge correlation are involved.
At high temperatures in the metallic phase, 
diffuse X-ray  spots are observed 
at $(\frac{2}{3},k,\frac{1}{4})$ and 
$(\frac{1}{3},k,\frac{3}{4})$, which suggests the presence 
of short-range $3\times 4$ charge correlation (figure \ref{fig4}) 
\cite{Watanabe2004,Watanabe2005}. 
Some anomalies have also been observed 
in the nuclear magnetic resonance 
experiments in a similar temperature range \cite{Miyagawa,Chiba}.
At around 200 K, 
the system undergoes a metal-insulator transition, 
accompanied by a structural phase transition to the so-called 
$\theta_d$ phase \cite{Seo}, 
in which the unit cell is doubled in the $c$ direction
\cite{HMori2,Watanabe2004,Watanabe2005,Miyagawa,Chiba,HMori3,Tajima,Yamamoto}.
In the $\theta_d$ phase, X-ray diffraction measurements have revealed that 
long-range horizontal stripe charge ordering 
with the modulation wave vector $(0,0,\frac{1}{2})$ 
takes place \cite{HMori1,HMori2,Watanabe2004,Watanabe2005}.

Theoretically, the origin of these charge correlations has been 
an issue of great interest \cite{Seorev}. 
A number of analytical and numerical studies have been performed 
using the extended Hubbard model (coupled with the lattice in some cases) 
described in the following section.
In the present paper, we review theoretical studies on the charge 
correlations in $\theta$-(BEDT-TTF)$_2X$.

\section{Theoretical model}

\begin{figure}
\begin{center}
\includegraphics[width=8cm]{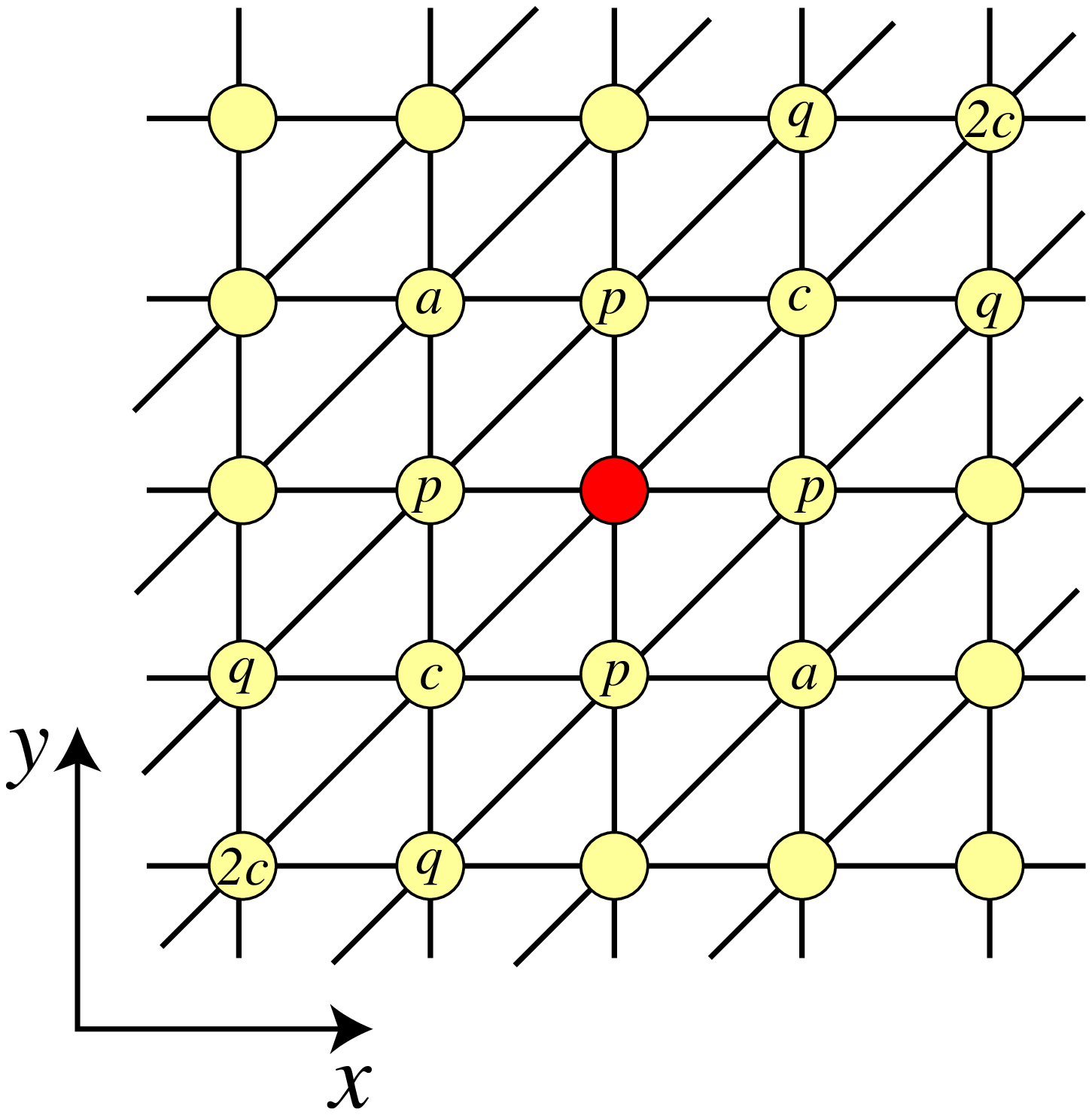}
\end{center}
\caption{Effective lattice structure using the $x$-$y$ 
unit cell. We specify the hoppings and the interactions by 
$c$, $p$, $a$, $q$, and $2c$, which specify the positions relative to the 
center (red) site.}
\label{fig6}
\end{figure}

\begin{figure}
\begin{center}
\includegraphics[width=8cm]{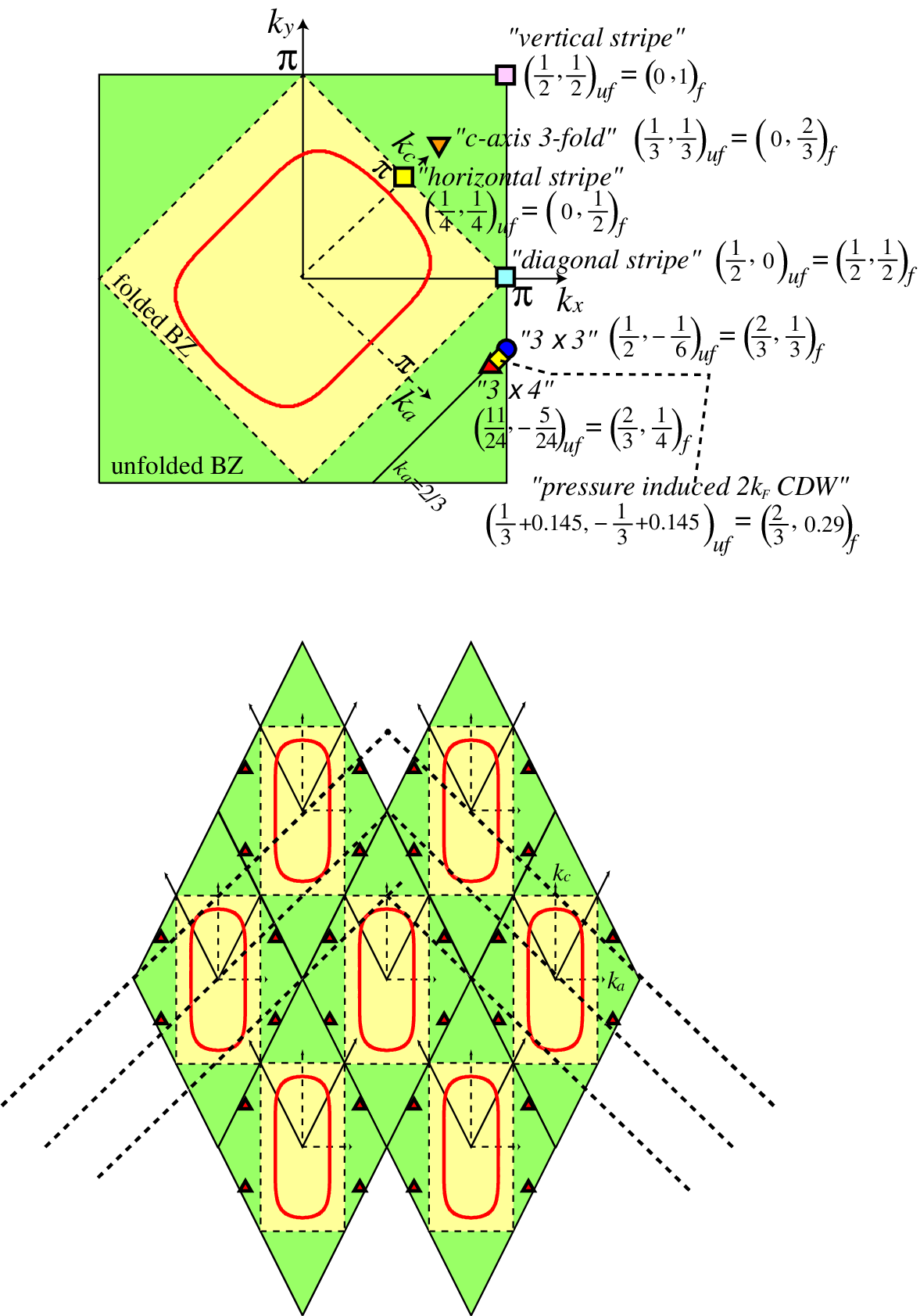}
\end{center}
\caption{Upper panel: relation between the folded and unfolded 
Brillouin zone. Modulation wave vectors corresponding to 
various correlations are shown.
The wave vectors are 
presented in units of the reciprocal lattice primitive vectors.
The red line shows the Fermi surface for $t_c=-0.3$. 
Lower panel: Brillouin zone in the upper panel 
squeezed in the $k_x=-k_y$ direction, and then placed repeatedly.
The dashed lines represent the diffuse sheets observed at room temperature
and below.}
\label{fig7}
\end{figure}

In this section, we describe an 
electronic model of $\theta$-(BEDT-TTF)$_2X$, where 
each molecule is considered as a ``site'' in the tight binding 
approximation. 
In the lattice structure of $\theta$-(BEDT-TTF)$_2X$ shown in 
figure \ref{fig2}, the direction of the molecules alternates along 
the $a$ axis resulting in the rectangular $a$-$c$ unit cell, 
but this alternation is irrelevant as far as the 
hopping integrals in the tight binding model are concerned.
Therefore, we can consider a unit cell ($x$-$y$)  
which is half the size of the 
usual unit cell, thereby obtaining the effective lattice structure 
shown in figure \ref{fig6}.
The corresponding Brillouin zone  becomes unfolded, and the 
relation between the unfolded and folded Brillouin zone is shown 
in figure \ref{fig7}. Here, the wave vectors are given in units of the 
reciprocal lattice primitive vectors, and the subscripts $uf$ and $f$
indicate whether the reciprocal vectors are those of the unfolded or  
folded Brillouin zone, respectively.
The wave vector of the experimentally observed 
horizontal stripe charge ordering 
is $(1/4,1/4)_{uf}$ in the unfolded Brillouin zone. 
The wave vector $(\frac{2}{3},k,\frac{1}{3}\sim\frac{1}{4})_{f}$ at 
which the diffuse X-ray spots are observed in the high temperature 
regime lie on the diagonal line labeled as 
$k_a=\frac{2}{3}$, which satisfies $k_x-k_y=\frac{2}{3}$ 
in the unfolded Brillouin zone. 
 Since these wave vectors correspond to 
charge correlation with threefold periodicity in the 
$a$-axis direction (figure \ref{fig4}), we will call these 
the ``$a$-axis threefold'' charge correlation.


Important progress in understanding the origin of the charge correlations 
in $\theta$-(BEDT-TTF)$_2X$ was made by Seo, who included the nearest 
neighbor off-site repulsions $V_p$ and $V_c$ 
in the model Hamiltonian (see figure \ref{fig2})\cite{Seo} in addition to the 
on-site interaction $U$, which had already been considered in the seminal 
studies of organic conductors 
by Kino and Fukuyama \cite{KF} and Seo and Fukuyama \cite{SeoFukuyama}.
Mori estimated the interaction parameters by directly calculating the 
Coulomb interaction using the highest occupied molecular orbitals of 
BEDT-TTF, neglecting the screening effect \cite{TMori}. 
This estimation shows that $V_p\sim V_c$ ($V_p$ is 
slightly smaller than $V_c$), 
while interactions $V_a$, $V_p$, and $V_{2c}$ are also 
of significant size compared with $V_p$ and $V_c$.

A number of theoretical studies 
on the charge correlations in  $\theta$-(BEDT-TTF)$_2X$  
have been performed using the extended Hubbard model 
on the lattice shown in figure \ref{fig6} \cite{Hanasaki}, 
where the Hamiltonian is given in the form 
\begin{equation}
H=\sum_{<i,j>,\sigma} 
t_{ij}c^{\dagger}_{i\sigma}c_{j\sigma}
+U\sum_{i}n_{i\uparrow}n_{i\downarrow}
+ \sum_{<i,j>}V_{ij} n_{i}n_{j},
\end{equation}
where $c^{\dagger}_{i\sigma}$ creates an electron 
with spin $\sigma = \uparrow, \downarrow$ at site $i$, 
and $n_{i\sigma}=c_{i\sigma}^\dagger c_{i\sigma}$.
The band filling, i.e., 
the average number of electrons per site (molecule), is 
fixed at $n=1.5$ in accord with the actual material.
In figure \ref{fig6}, the letters $p$, $c$, $a$, $\cdots$ denote the 
relative positions with respect to the red site at the center, 
and we use these letters to specify the range of the hopping integrals and 
the off-site repulsive interactions.
For the kinetic energy terms, we consider hoppings  
$t_p$ and $t_c$, where $t_p\ (\simeq 0.1\; {\rm eV})$ 
is taken as the unit of energy. 
$X=$Cs$M'$(SCN)$_4$ ($M'=$Co,\ Zn) has small values of $t_c$, while 
$X=$Rb$M'$(SCN)$_4$ ($M'=$Co,\ Zn) and $X=$CsCo(SCN)$_4$ under pressure
have relatively large values of $t_c\sim -0.3$ to $-0.4$ 
\cite{Watanabe1999}.

\section{Stripe-type charge correlations}
Three types of stripe-type charge correlations,  
vertical, horizontal, and diagonal (figure \ref{fig3}),  
were proposed by Seo \cite{Seo}. The wave vectors of the three 
patterns correspond to 
$(1/2,1/2)_{uf}=(0,1)_f$, $(1/4,1/4)_{uf}=(0,1/2)_f$, 
and $(1/2,0)_{uf}=(1/2,1/2)_f$, respectively. 
Intuitively, the horizontal and diagonal stripe 
states are favored when $V_p$ is small, while the vertical stripe 
state is favored when $V_c$ is small, as can be understood from 
figure \ref{fig3}. In ref. \cite{Seo}, 
the energies of various types of charge correlation patterns 
were calculated within the mean field approximation.
The calculation shows that the diagonal stripe is more stable than 
the horizontal and vertical stripe states 
when $V_p$ is small, while the vertical stripe state is 
stable for small $V_c$.
Exact diagonalization studies on these stripe orders 
have also been performed \cite{Clay,Seo2}.

\begin{figure}
\begin{center}
\includegraphics[width=8cm]{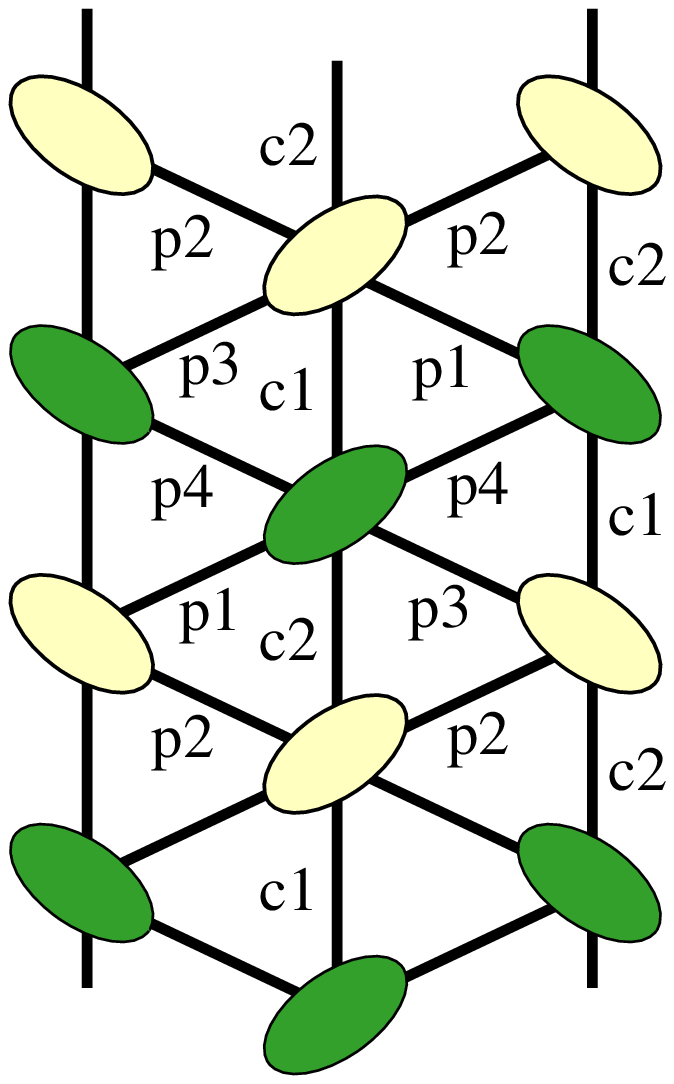}
\end{center}
\caption{$\theta_d$ lattice. The hopping integrals are 
$c1=1.5$, $c2=5.2$, $p1=16.9$, $p2=-6.5$, $p3=2.2$, and $p4=-12.3$ in 
units of $10^{-2}$ eV \cite{Watanabe2004}. The green molecules are the 
charge rich ones in the horizontal stripe state.}
\label{fig8}
\end{figure}

Seo further showed that the horizontal stripe state is much 
stabler in the $\theta_d$ 
phase of the crystal structure than in the $\theta$ phase, where 
the hopping integrals are modulated as shown in figure \ref{fig8}
\cite{commenttheta_d} In this $\theta_d$ lattice structure, one-dimensional 
chains of molecules connected by the large hopping $p4$ are weakly  
coupled by other small hoppings. In Seo's calculation, 
the competition between the diagonal and  horizontal stripe 
states was subtle, but 
the resulting effective spin Hamiltonian in the horizontal stripe phase is 
a quasi-one-dimensional Heisenberg chain along the $p4$ hopping integral,  
which explains the Bonner-Fisher type temperature dependence of the 
magnetic susceptibility \cite{HMori1,HMori2}. 
From this viewpoint, Seo concluded that the horizontal stripe ordering 
is the most probable state in the insulating phase of 
$X=$Rb$M'$(SCN)$_4$. This was later confirmed by 
X-ray experiments \cite{Watanabe2004,Watanabe2005} 
as mentioned in the introduction.

\section{Competition between stripe and threefold charge correlations}

\subsection{$c$-axis threefold state}
After the seminal study by Seo \cite{Seo}, 
Mori performed an analysis in the static limit (where the band structure 
is neglected), and pointed out the possibility of 
the nonstripe-type charge correlation 
shown in figure \ref{fig3} \cite{TMori},
 which has threefold periodicity in the 
$c$-axis direction. We will call this the ``$c$-axis threefold'' charge 
correlation. The corresponding wave vector is $(1/3,1/3)_{uf}=(0,2/3)_f$.
In real space, the $c$-axis threefold state is favored when 
$V_p\simeq V_c$ and $V_a=V_q=V_{2c}=0$ 
because this configuration minimizes the energy cost of $V_p$ and 
$V_c$, as can be seen from figures \ref{fig2} 
and \ref{fig3}. 

Kaneko and Ogata 
extended Seo's mean field study to take into account the possibility of 
this $c$-axis threefold state \cite{KanekoOgata}.
They found that this state has 
a lower energy than the vertical, diagonal, and horizontal stripe states 
when $V_p\sim V_c$ and  $V_a=V_q=V_{2c}=0$.   
Hotta and coworkers 
used exact diagonalization for a spinless half-filled 
model, which can be considered as an effective model of $\theta$-type 
compounds in the large on-site $U$ limit, and they showed the 
presence of $c$-axis 
threefold charge correlation for $V_p\sim V_c$ \cite{Hotta,Hotta2}.

\begin{figure}
\begin{center}
\includegraphics[width=8cm]{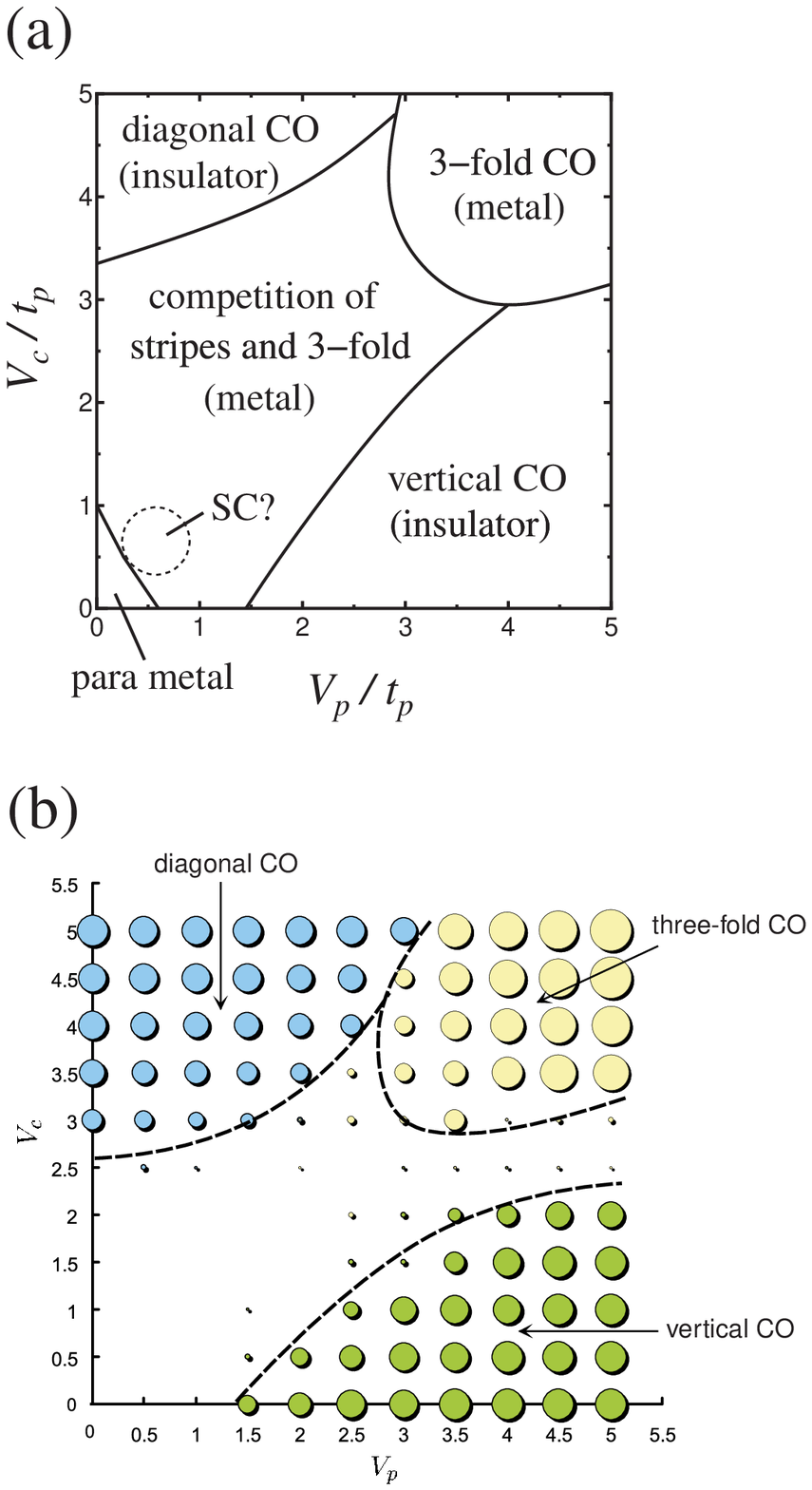}
\end{center}
\caption{Ground-state phase diagram obtained by (a) variational 
Monte Carlo and (b) density matrix renormalization group for 
$t_c=0$, $U=10$, and $V_a=V_q=V_{2c}=0$. In (b), the diameter of the circles 
is proportional to the hole density at the hole-rich sites. 
(a) Reproduced with permission from ref. \cite{WataOgata}. Copyright 2006 
by the Physical Society of Japan. (b) Reproduced with permission from 
ref. \cite{Nishimoto}. Copyright 2008 by the American Physical Society.}
\label{fig9}
\end{figure}

A variational Monte Carlo  study by 
Watanabe and Ogata \cite{WataOgata}, 
and a later density matrix renormalization 
group study by Nishimoto {\it et al.} \cite{Nishimoto} 
revealed the phase diagram of the model  in $V_p$-$V_c$ space with $t_c=0$
and $U/t_p=10$. In the variational Monte Carlo study, 
the energy of each ordered state 
was calculated and compared, while in the density matrix renormalization group 
study, the real-space charge modulation  
was directly observed.
As shown in figure \ref{fig9}, the two approaches give similar results, 
where the $c$-axis threefold state appears in a regime with $V_p\sim V_c$ 
between the diagonal ($V_p < V_c$) 
and vertical stripe ($V_p > V_c$) regimes.

A random phase approximation (RPA) 
analysis gives a clear view, in the reciprocal space, 
as to why the $c$-axis three fold state
is favored in the $V_p\sim V_c$ regime, 
and also why the diagonal stripe is more favored than 
the horizontal stripe \cite{Kurokitheta}.
Within the RPA \cite{KobaOgata,Scalapino,TanakaOgata}, 
the charge susceptibility $\chi_{c}$ is given as 
\begin{eqnarray}
\label{4}
\chi_{c}(\Vec{q})=\frac{\chi_{0}(\Vec{q})}
{1 + (U + 2V(\Vec{q}) )\chi_{0}(\Vec{q})}.
\label{chargeRPA}
\end{eqnarray}
Here $\chi_{0}$ is the bare susceptibility given by 
\begin{equation}
\chi_{0}(\Vec{q})
=\frac{1}{N}\sum_{\Vec{p}} 
\frac{ f(\epsilon_{\Vec{p +q}})-f(\epsilon_{\Vec{p}}) }
{\epsilon_{\Vec{p}} -\epsilon_{\Vec{p+q}}},
\end{equation}
with
$\varepsilon_{\Vec{k}}$ being the band dispersion given as
\begin{equation}
\varepsilon_{\Vec{k}}=2t_p[\cos(k_x)+\cos(k_y)]+2t_c\cos(k_x+k_y)
\end{equation}
and 
$f(\epsilon_{\Vec{p}})=1/(\exp(\epsilon_{\Vec{p}}-\mu)/T) + 1)$ is 
the Fermi distribution function. 
When the nesting of the Fermi surface is good, 
$\chi_0(\Vec{q})$ is maximized at the nesting vector.
$V(\Vec{q})$ is the Fourier transform of the off-site repulsions, 
given as
\begin{eqnarray}
V(\Vec{q})&=&2V_p[\cos(q_x)+\cos(q_y)] \nonumber\\
&+&2V_c\cos(q_x+q_y)
+2V_a\cos(q_x-q_y)\nonumber\\
&+&2V_q[\cos(2q_x+q_y)+\cos(q_x+2q_y)]\nonumber\\
&+&2V_{2c}\cos(2q_x+2q_y)
\label{3}
\end{eqnarray}

\begin{figure}
\begin{center}
\includegraphics[width=5cm]{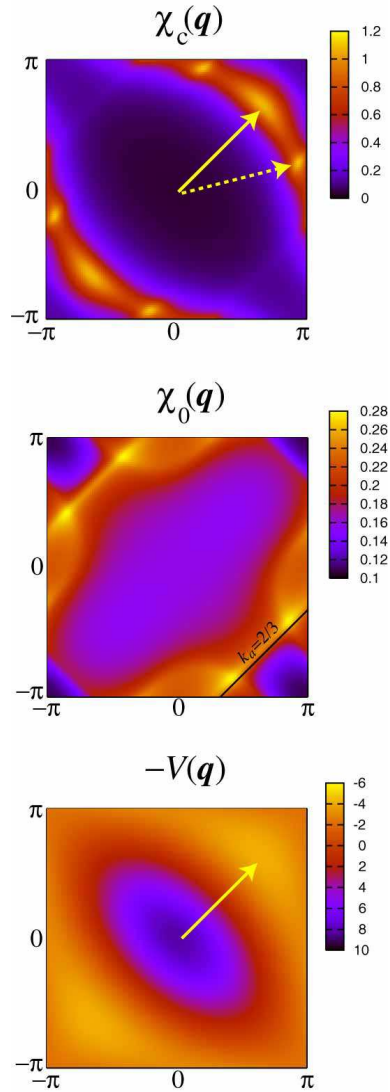}
\end{center}
\caption{
Contour plots of 
the RPA result of the charge susceptibility $\chi_c$,
the bare susceptibility $\chi_0$, and the Fourier transform of the 
off-site interactions $-V(\Vec{q})$. $t_c=-0.3$, 
$U=3$, $V_p=1.3, V_c=1.5$, 
$V_a=V_q=V_{2c}=0$, and $T=0.05$. 
The solid (dashed) arrow in $\chi_c(\Vec{q})$ 
indicates the dominant (subdominant) peak at wave vectors 
close to $(1/3,1/3)_{uf}$ ($(1/2,0)_{uf}$). The solid arrow 
in $-V(\Vec{q})$ is the wave vector $(1/3,1/3)_{uf}$ \cite{Kurokitheta}.
}
\label{fig10}
\end{figure}

In figure \ref{fig10}, the RPA charge susceptibility 
for $t_c=-0.3$, $U=3$, $V_c=1.5$, $V_p=1.3$, and $V_a=V_q=V_{2c}=0$ is shown. 
This choice of $t_c$ 
corresponds to the case of $X=$Rb$M'$(SCN)$_4$ ($M'$=Zn, Co) or 
$X=$CsCo(SCN)$_4$ under pressure. 
One notices that there is a peak 
at $(1/3,1/3)_{uf}$, which corresponds to 
$(0,2/3)_f$ in the original Brillouin zone, 
namely, the modulation wave vector of the 
$c$-axis threefold correlation. This is due to a peak 
in $-V(\Vec{q})$ near $\Vec{q}=(1/3,1/3)_{uf}$ as shown 
in figure \ref{fig10}. Namely, although $\chi_0(\Vec{q})$ is 
maximized around $(2/3,1/4)_{f}=(11/24,-5/24)_{uf}$,  the effect of 
 $-V(\Vec{q})$ is strong in the denominator of the RPA 
charge susceptibility eq.(\ref{chargeRPA}),
thereby minimizing the denominator and maximizing $\chi_c(\Vec{q})$ at 
$(1/3,1/3)_{uf}$.
In addition to the $c$-axis threefold peak, 
a subdominant peak also appears around the wave vector close to 
$(1/2,0)_{uf}$, i.e., the wave vector of the 
diagonal stripe. 
This is due to a combination of two effects: 
(i) $\chi_0(\Vec{q})$ takes large values in the region extending 
from the nesting vector position toward $(1/2,0)_{uf}$,  and 
(ii) the region where $-V(\Vec{q})$  takes large values tends to extend
toward $(1/2,0)_{uf}$, particularly when $V_p<V_c$.
A similar result is also obtained for smaller values of $t_c$, 
although the subdominant peak near $(1/2,0)_{uf}$   
becomes relatively weaker.
Such a two-peak structure has also been found in the 
dynamical charge correlation function calculated using exact diagonalization, 
where peaks at low energies are found 
at $(1/2,0)_{uf}$ and $(1/3,1/3)_{uf}$ \cite{Nishimoto}.

Udagawa and Motome \cite{Udagawa} further 
investigated the temperature 
dependence of these peak structures in the charge susceptibility, 
and showed that 
the $(1/3,1/3)_{uf}$ peak is nearly temperature-independent, 
while the peak near $(1/2,0)_{uf}$ increases in magnitude 
as the temperature is lowered. This is because 
the $(1/3,1/3)_{uf}$ peak is mainly due to the 
nearest-neighbor electron-electron interaction, while the 
$(1/2,0)_{uf}$ peak is related to the Fermi surface 
nesting, which becomes more effective at low temperatures.
From this temperature dependence, 
they claimed that these peaks correspond 
to the two diffuse X-ray spots observed in 
$X=$Cs$M'$(SCN)$_4$ ($M'=$Co,\ Zn) 
\cite{Nogami,Watanabe1999} (figure \ref{fig5}),  
although the wave vectors are different.

\subsection{Effect of electron-lattice couplings}

As seen in the previous section, a purely electronic model 
on the $\theta$-type lattice fails to reproduce the 
horizontal stripe charge correlation observed 
experimentally. 
Following the line of Seo's study \cite{Seo},
Kaneko and Ogata \cite{KanekoOgata} performed a mean-field analysis 
on the $\theta_d$ lattice structure, where competition between 
the $c$-axis threefold state and the horizontal stripe was found to 
take place.

Tanaka and Yonemitsu took into account Peierls-type 
electron-lattice couplings,  
and performed a Hartree-Fock calculation \cite{TanakaYone}. 
They showed that 
the horizontal stripe state is stabilized by 
a self-consistently determined lattice distortion into the $\theta_d$ phase
when $V_p/V_c<1$. The stabilization of the horizontal stripe 
state by electron-lattice coupling has been further supported by 
an exact diagonalization study by Miyashita and Yonemitsu \cite{Miyashita}.

Udagawa and Motome included the Su-Schrieffer-Heeger type
electron-phonon interaction, which also takes into account the 
effect of the lattice distortion, and performed a Hartree-Fock 
calculation \cite{Udagawa}. They found that the horizontal stripe state 
is stable for sufficiently large electron-phonon couplings. 

Note that 
although these calculations succeed in reproducing the 
horizontal stripe charge correlation, 
the competing candidates considered therein  
do not include the $a$-axis threefold correlation observed experimentally.

\subsection{$a$-axis threefold charge correlation}

\begin{figure}
\begin{center}
\includegraphics[width=5cm]{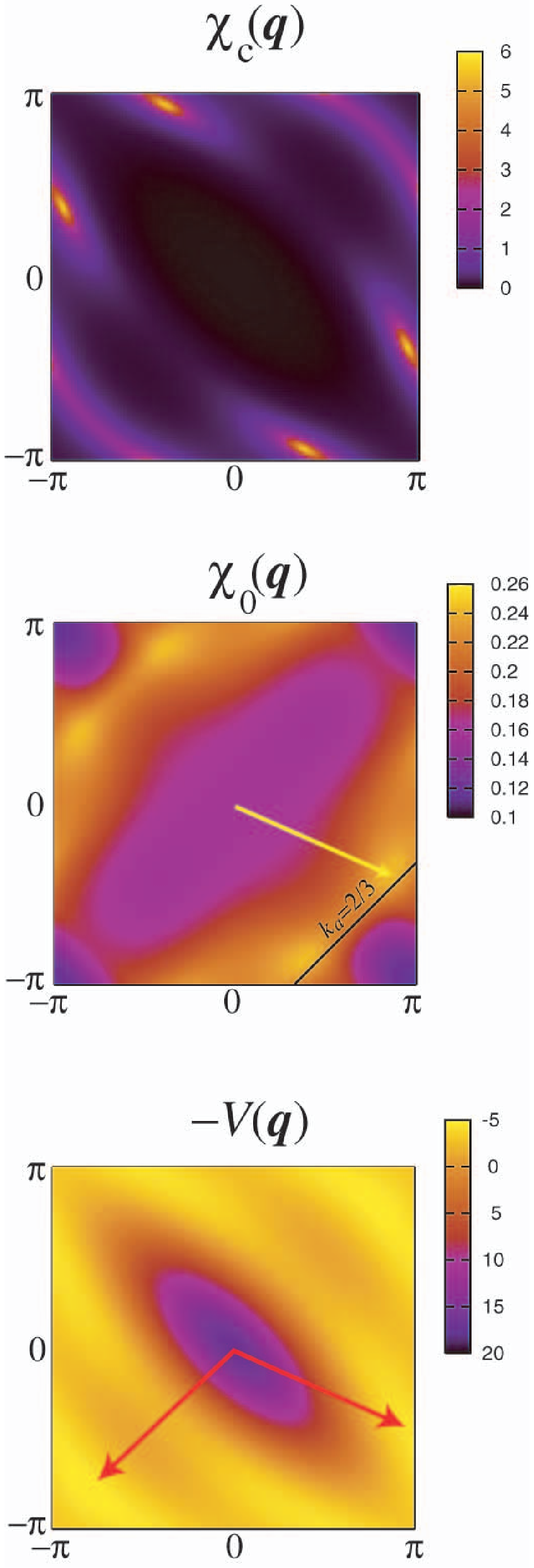}
\end{center}
\caption{
Contour plots of 
the RPA result for the charge susceptibility $\chi_c$,
the bare susceptibility $\chi_0$, and the Fourier transform of the 
off-site interactions $-V(\Vec{q})$. $t_c=-0.3$, 
$U=4$, $V_p=1.8, V_c=2.0$, 
$V_a=0.4$,$V_q=0.7$, $V_{2c}=1.1$, and $T=0.25$. 
The arrow in $\chi_0(\Vec{q})$ 
indicates the peak position, i.e., the 
nesting vector of the Fermi surface.
The arrows in $-V(q)$ indicate the area at which $-V(q)$ is 
broadly maximized \cite{Kurokitheta}.
}
\label{fig11}
\end{figure}

\begin{figure}
\begin{center}
\includegraphics[width=5cm]{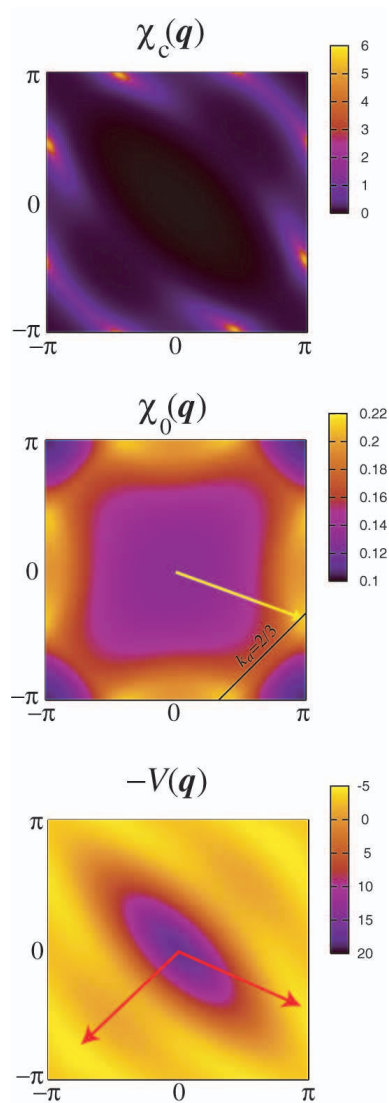}
\end{center}
\caption{ Plots similar to figure \ref{fig11} except that 
$t_c=-0.05$ and $T=0.15$ \cite{Kurokitheta}.}
\label{fig12}
\end{figure}

\begin{figure}
\begin{center}
\includegraphics[width=5cm]{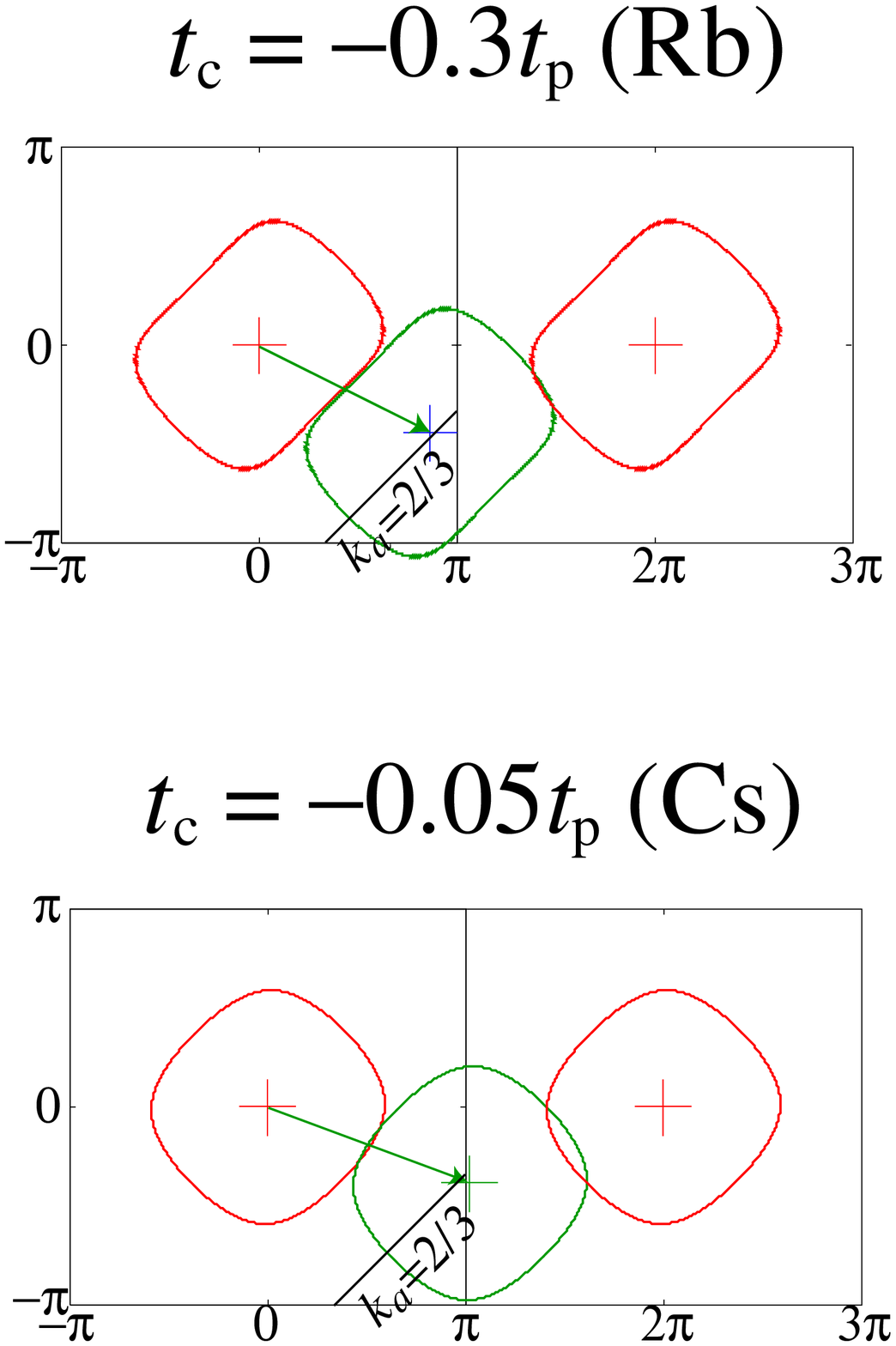}
\end{center}
\caption{Comparison of the Fermi surface nesting between 
the cases of $t_c=-0.3$ and $-0.05$.}
\label{fig13}
\end{figure}

As seen in the previous subsections, 
models that consider only $U$, $V_p$, and $V_c$ as 
electron-electron repulsions have a tendency toward 
the $c$-axis threefold state formation 
in the realistic parameter regime of $V_p\sim V_c$. 
This state competes with the diagonal (horizontal) correlation in the 
absence (presence) of electron-lattice coupling \cite{Hottacomment2}.
Correspondence between the $c$-axis threefold charge correlation 
and the experimentally observed 
short range $a$-axis threefold charge correlation 
has often been discussed, but 
the wave vectors of the two threefold correlations are different.

Kuroki performed an RPA analysis on a model that considers the 
distant interactions $V_a$, $V_q$, and $V_{2c}$ 
in order to investigate the origin of the 
$a$-axis threefold correlation \cite{Kurokitheta}. 
In figure \ref{fig11}, 
the charge susceptibility for $t_c=-0.3$ is shown, where 
a peak is found near the $a$-axis threefold correlation wave vectors.
Here only the result for certain values of $V_a$, $V_q$, and $V_{2c}$ is 
shown, but in ref. \cite{Kurokitheta} it was found that similar results 
are obtained within a certain range of $V_a$, $V_q$, and $V_{2c}$ 
values that are smaller  than $\sim V_p/2$ and $\sim V_c/2$.
This result can be understood as follows. In the presence of the 
distant interactions, $-V(\Vec{q})$ is now broadly 
maximized in a region that includes the nesting vector position, as shown in 
figure \ref{fig11}. 
Therefore, the denominator in eq.(\ref{chargeRPA}) is now 
minimized at a position close to the nesting vector, resulting in 
a charge susceptibility peak position that nearly 
coincides with the diffuse X-ray spot positions. 
As mentioned in the introduction, the relevance of the 
Fermi surface nesting to the $a$-axis threefold correlation 
had already been pointed out in ref. \cite{Watanabe1999}. 
Thus, the $a$-axis threefold charge correlation, within this scenario,  
should be considered as being conceptually  
different from the $c$-axis threefold charge ordering 
in that it is not related to the Fermi surface.

\begin{figure}
\begin{center}
\includegraphics[width=8cm]{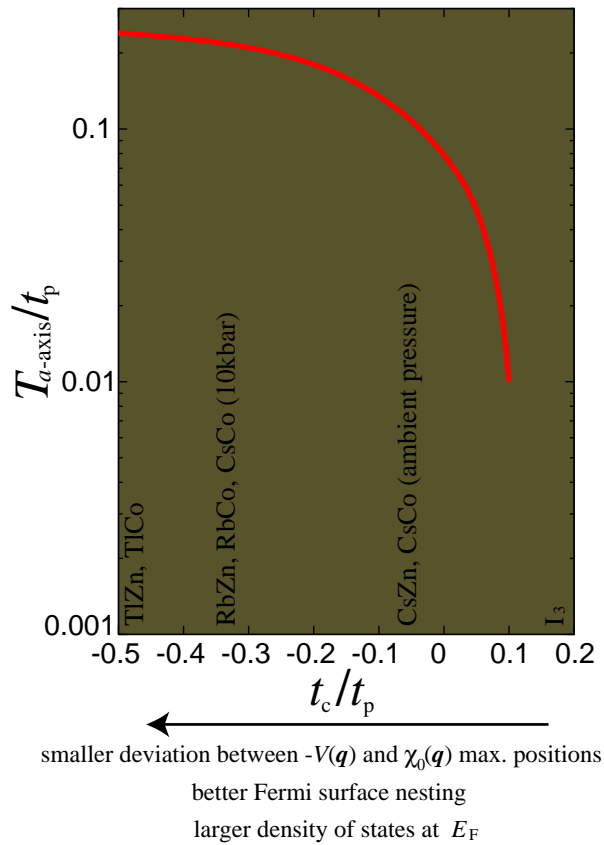}
\end{center}
\caption{
$T_{a-{\rm axis}}$ plotted as a function of $t_c$ for 
$U=4$, $V_p=1.8$, $V_c=2.0$, $V_a=0.4$, $V_q=0.7$, and $V_{2c}=1.1$
\cite{Kurokitheta}.
$t_p\simeq 0.1 {\rm eV}\sim 1000$ K is taken as the unit of temperature.
The values of $t_c/t_p$ for $X=MM'$(SCN)$_4$ are taken from 
table II in ref. \cite{Watanabe1999}. For $X=$I$_3$, 
the value of $t_p$ is taken from ref. \cite{HMori2}, while 
that of $t_c$ is from ref. \cite{HMoriPv}.}
\label{fig14}
\end{figure}

In figure \ref{fig12}, the result for $t_c=-0.05$ with 
the other parameters having the same values as those in 
figure \ref{fig11} is shown. 
This choice of $t_c$ 
corresponds to $X=$Cs$M'$(SCN)$_4$ at ambient pressure \cite{Watanabe1999}.
In this case, the peak position of the charge susceptibility 
moves toward the unfolded Brillouin zone edge, i.e., closer to 
$(\frac{1}{2},-\frac{1}{6})_{uf}=(\frac{2}{3},\frac{1}{3})_f$, 
compared with the case when $t_c=-0.3$. 
This is because the nesting vector 
is closer to $(\frac{2}{3},\frac{1}{3})_f$, 
as can be seen in the figure of $\chi_0$, which in turn can be 
understood from the Fermi surface shown in figure \ref{fig13}. 
This tendency 
appears to be consistent with the experimental finding that 
diffuse X-ray spots at high temperatures 
are observed at $q_c=\frac{1}{3}$  
in $X=$Cs$M'$(SCN)$_4$ at ambient pressure, while 
they are at $\frac{1}{4}\leq q_c \leq \frac{1}{3}$ in 
$X=$Cs$M'$(SCN)$_4$ under  pressure and in $X=$Rb$M'$(SCN)$_4$.

In figure \ref{fig14} the temperature $T_{a-{\rm axis}}$ 
at which the RPA charge susceptibility diverges at a certain $\Vec{q}$  
is shown as a function of $t_c$ \cite{Kurokitheta}. 
$T_{a-{\rm axis}}$ decreases with increasing 
$t_c$ (from $t_c<0$ to $t_c>0$), 
and sharply drops when $t_c>0$. In fact, the increase in $t_c$ induces 
several effects. First, the nesting of the Fermi surface is 
degraded (see figure \ref{fig13}),   
second, the nesting vector deviates from the region where 
$-V(\Vec{q})$ is maximized, and third, 
the density of states near the Fermi level becomes smaller.
These factors together deteriorate the charge ordering.
The decrease in the density of states with 
increasing $t_c$ has also been pointed out in ref. \cite{Hottacomment}.
Figure \ref{fig14} is reminiscent of the 
phase diagram of the $\theta$-(BEDT-TTF) family 
obtained by Mori {\it et al}. \cite{HMori2}, where charge ordering 
takes place 
for compounds having large negative $t_c$ such as $X=$Rb$M'$(SCN)$_4$, 
while it does not occur for compounds having small 
or positive $t_c$ \cite{HMoriPv,HKobayashi,Tamura2} such as $X=$I$_3$. 
However, note that in figure \ref{fig14}, 
the charge susceptibility diverges 
at a modulation wave vector close to 
the nesting vector of the Fermi surface 
(around $(\frac{2}{3},\frac{1}{3}-\frac{1}{4})_f$ 
for $t_c<0$), while the charge ordering temperature 
in the experimental phase diagram is that for the 
$(0,k,\frac{1}{2})_f$ ordering, i.e., the horizontal stripe state.
The resemblance between the two figures suggests a 
relationship between the horizontal stripe and the $a$-axis threefold 
correlations.
This may also be consistent with an experimental 
analysis under uniaxial pressure, which concluded that 
the value of $t_c$ is the key factor that dominates the occurrence of 
horizontal charge ordering \cite{Kagoshima2}.
A possible relation between the $a$-axis threefold and the 
horizontal stripe orderings will be mentioned in the  next subsection.

\subsection{Future problems regarding the competition}
\label{future}
Since the distant electron-electron interactions $V_a$, $V_q$, 
and $V_{2c}$ are found to be necessary to reproduce 
the experimentally observed $a$-axis threefold charge correlations, 
it is reasonable to assume that the related experimental observations 
have to be understood within such a model. 

\begin{figure}
\begin{center}
\includegraphics[width=8cm]{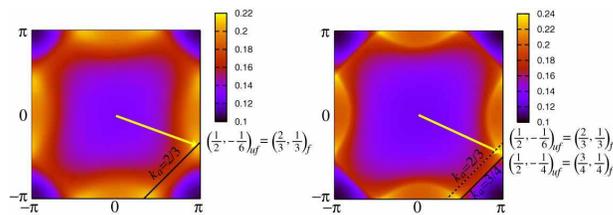}
\end{center}
\caption{$\chi_0(\Vec{q})$ for $t_c=-0.05$ at $T=0.15$ ($\sim 150$ K)  
(left) and at $T=0.02$ ($\sim 20$ K) (right).}
\label{fig15}
\end{figure}

One problem to be resolved is the temperature dependences of the two diffuse 
X-ray peaks observed in $X=$CsCo(SCN)$_4$ \cite{Watanabe1999}.  
Since the peak in the charge susceptibility partially originates 
from the Fermi surface nesting in the scenario described in the 
previous subsection, the maximum value of the peak 
increases with decreasing temperature, which is inconsistent with 
the experimental observation that the X-ray peak corresponding to the 
$a$-axis threefold charge fluctuation is nearly temperature-independent 
below 90K \cite{Watanabe1999}. Regarding this problem, 
it is possible that the anions, which are 
not taken into account in previously studied models, may play some role. 
Namely, in the X-ray scattering experiment, 
diffuse sheets (or planes) along the $2a+c$ and the $2a-c$ directions 
are observed at room temperature below, which is considered to be 
due to the effect of anions 
(the diffuse sheets are shown in the lower panels of 
figure \ref{fig5} and figure \ref{fig7}) \cite{Nogami}. Actually, 
the wave vectors $(\frac{2}{3},k,\frac{1}{3})$ ($a$-axis three fold) 
and $(0,k,\frac{1}{2})$ (horizontal stripe) are on these diffuse 
sheets, and they ``grow'' from these 
sheets as the temperature is lowered, 
suggesting the relevance of the anions.
Kuroki considered the effect of these diffuse sheets phenomenologically 
by adding a term that enhances the 
charge susceptibility along the line $2a+c$ 
to the denominator of eq.(\ref{chargeRPA}), and he found that 
the maximum value of the charge susceptibility saturates 
at low temperatures \cite{KKunpub}. This is because the wave vector at which 
$\chi_0(\Vec{q})$ takes its maximum 
gradually moves away from $(2/3,1/3)_f$ 
toward $(3/4,1/4)_f$  as the 
temperature is lowered as shown in figure \ref{fig15}  
(note that the nesting is less good 
for the case of $X=$CsCo(SCN)$_4$), so that the 
cooperation between the diffuse sheets and the Fermi surface nesting 
becomes degraded. 

Another problem is the origin of the 
horizontal stripe charge correlation. As mentioned in the previous sections, 
in the absence of the 
distant interactions $V_a$, $V_q$, and $V_{2c}$, 
the horizontal stripe state is shown to be favored 
by electron-lattice coupling that distorts the lattice structure 
into the $\theta_d$-type. Nevertheless, 
it remains a future 
problem as to whether this conclusion holds in the 
presence of the distant interactions. 
Moreover, even if the origin of the horizontal stripe in the 
Rb salt is the lattice distortion, the development of the 
horizontal stripe correlation at low temperatures in the Cs salt, 
where such a distortion has not been observed experimentally, remains puzzling.
Related to this point, Kuroki pointed out the possibility 
that the horizontal stripe charge correlation arises 
from the $a$-axis threefold correlation \cite{Kurokitheta}. 
For the case of the 
Cs salt, although the nesting vector is close to $(2/3,1/3)_f$ 
at high temperatures, it moves toward $(3/4,1/4)_f$  
(note that we are adopting the 
{\it folded} Brillouin zone scheme here) upon 
lowering the temperature as mentioned above. 
Thus, at low temperatures, 
we have two different nesting vectors close to 
$(3/4,1/4)_f$ and $(-3/4,1/4)_f$.
Their coexistence in the momentum space 
might result in a modulation having the wave vector 
$(0,1/2)_f=(3/4,1/4)+(-3/4,1/4)$, which corresponds to that of the 
horizontal stripe.
 This may be considered as a 
kind of ``$4k_F$'' correlation in that the wave vectors 
$(3/4,1/4)_f$ and $(-3/4,1/4)_f$ are close to the nesting vectors of the 
Fermi surface, namely, the ``$2k_F$'' wave vectors at low temperatures. 
Investigation of the 
occurrence of the horizontal charge correlation due to 
the interplay between such a ``$4k_F$'' correlation and the electron-lattice 
couplings (including the interaction with anions) will serve as an interesting 
future area of research.

There are several studies in which the pairing 
symmetry of superconductivity in $X=$I$_3$ has been discussed 
\cite{WataOgata,Kurokitheta,TanakaOgata,Merino}. Since 
this superconductivity lies in the vicinity of the 
charge-ordering phase in Mori's electronic phase diagram (figure \ref{fig1}),
one may expect the occurrence of unconventional superconductivity 
due to charge fluctuations. In such a case, the form of the 
gap function strongly depends on the wave vector dependence of the 
charge fluctuation.
For instance, in a model that considers electron-electron 
interactions up to nearest neighbors, the coexistence of 
spin and charge fluctuations results in competition between
spin-singlet $d_{xy}$-wave-like pairing and triplet $f$-wave-like pairing 
\cite{WataOgata,TanakaOgata}.
On the other hand, in the model that takes into account the 
next nearest neighbor interaction, the gap function can be 
more complicated \cite{Kurokitheta}.
Since it can be expected that 
the many body interactions do not depend strongly on the anions  
(although the band structure does),
one must discuss the pairing symmetry for $X=$I$_3$ using 
a Hamiltonian with appropriate interactions 
that correctly reproduces the competition between 
the $a$-axis threefold  and the horizontal stripe charge correlations 
in the case of Rb and Cs salts.  In this sense, the 
superconducting pairing symmetry 
should be theoretically determined after the origin of the charge 
correlations is completely understood.

\section{Conclusion} 

In the present paper, we have reviewed 
theoretical studies on the 
charge correlations in $\theta$-(BEDT-TTF)$_2MM'$(SCN)$_4$.
It now seems clear that within the purely electronic model on
the $\theta$-type lattice with the on-site $U$ and the  
nearest neighbor $V_p$ and $V_c$ interactions, the diagonal stripe, 
$c$-axis threefold, and vertical stripe charge correlations are 
favored in the regimes $V_p< V_c$, $V_p\sim V_c$, and $V_p> V_c$, 
respectively. In the realistic parameter regime of $V_p\sim V_c$, 
there is competition between the $c$-axis threefold state and the 
diagonal stripe state. Since these are different from the 
experimentally observed  $a$-axis threefold and 
the horizontal stripe charge correlations, 
additional effects have to be taken into account to 
understand the experimental results. The electron-lattice coupling, 
which tends to distort the lattice into the $\theta_d$-type,  
is found to favor the horizontal stripe state, 
suggesting that the occurrence of this stripe state 
in actual materials is not of purely electronic origin. 

On the other hand, the inclusion of the electron-lattice coupling does not 
explain the presence of the short-range $a$-axis threefold charge correlation,
and we have to add distant (next-nearest-neighbor) electron-electron 
repulsions to understand this phenomenon.
Related to this point, a 
transport experiment has indicated that 
the repulsive interaction in 
$\theta$-(BEDT-TTF)$_2M$Zn(SCN)$_4$ ($M=$Cs,Rb) is long-range \cite{Yamaguchi}.
In fact, the presence of such a distant interaction has also been 
pointed out theoretically \cite{KobayashiOgata,Suzumura,YK} 
for quasi-one-dimensional organic compounds 
(TMTSF)$_2$X in the context of coexisting $2k_F$ spin and $2k_F$ charge 
density waves \cite{PR,Kagoshima} and 
spin triplet pairing superconductivity 
\cite{Kurokirev,YK,KAA,KY,Fuseya,Nickel,Aizawa}.
Quite recently, Tahara {\it et al.} estimated the 
electron-electron interaction in $\kappa$-(BEDT-TTF)$_2X$ from 
a first principles calculation, 
and also obtained a long ranged interaction \cite{Tahara}.
Thus, it is possible that the long-range nature of the 
electron-electron interaction may be common in these organic materials.

So far there have been few theoretical studies that take into account the 
distant electron-electron interactions; thus, it is not clear whether 
competition between the $a$-axis threefold and horizontal 
stripe states takes place in the presence of electron-lattice 
couplings in such models. At the present stage, 
one cannot completely rule out the 
possibility that strong correlation 
effects that are not taken into account in RPA or mean-field 
analysis may result in the horizontal stripe state in the 
presence of distant interactions.
Also, as mentioned in section \ref{future},  
it is possible that the anions, which have not been taken into account in 
theoretical models so far, play some role in the 
occurrence of these charge correlations.
The minimal theoretical model may be more complicated than 
expected, and more theoretical study must be carried out 
to fully understand the physics of charge correlations  
in $\theta$-(BEDT-TTF)$_2X$.

\ack
The author would like to thank  H. Mori, T. Mori,  H. Seo, 
H. Fukuyama, Y. Nogami, M. Watanabe, T. Yamaguchi, 
Y. Ohta, M. Ogata, H. Watanabe, Y. Tanaka, C. Hotta, 
M. Udagawa, Y. Motome, M. Imada, K. Nakamura, R. Arita, and H. Aizawa 
for discussions. The author acknowledges 
Grants-in-Aid for Scientific Research from the Ministry of Education, 
Culture, Sports, Science and Technology of Japan, and from the Japan 
Society for the Promotion of Science.


\begin{thebibliography}{99}
\bibitem{commentorder} The term ``charge ordering'' is usually 
used in a sense that it is not related to the Fermi surface, while 
a ``charge density wave'' occurs as a result of Fermi surface nesting. 
As mentioned in the present article, the origin of the 
experimental observation regarding the charge correlation in 
$\theta$-(BEDT-TTF)$_2 X$ is still controversial, so that we cannot 
definitely state whether or not 
the Fermi surface is related to these phenomena 
at the present stage. Thus, in the present article, 
we mainly use the term ``charge correlation'' when we need to 
include various possibilities as the origin of these phenomena.
\bibitem{Hottacomment2} In ref. \cite{Hotta}, the horizontal stripe 
is found to be more competitive than the diagonal stripe. Since 
the spinless fermion model is adopted in this study, 
this may be the origin of the discrepancy.
\end{thebibliography}
\end{document}